\documentclass[aps,twocolumn,prd,preprintnumbers,amsmath,amssymb,superscriptaddress,nofootinbib]{revtex4-2}

\usepackage{graphicx} 
 \usepackage{color}
 \usepackage{epsfig}
 \usepackage{ulem}

\begin{document}

\title{
PeV Gravitino,  Weak-scale Higgsino and GeV Axino 
in KKLT
}

\author{Kyu Jung Bae} 
\email{kyujung.bae@knu.ac.kr}
\affiliation{Department of Physics, Kyungpook National University, Daegu 41566, Korea}
\author{Kwang Sik Jeong} 
\email{ksjeong@pusan.ac.kr}
\affiliation{Department of Physics, Pusan National University, Busan 46241, Korea}
\preprint{PNUTP-21-A12}

\begin{abstract}

We realize high scale supersymmetry in the mirage mediation. 
The Higgs sector is extended with the Peccei-Quinn symmetry,  and 
the higgsino mass term is generated by the Kim-Nilles mechanism.
In particular,  the Peccei-Quinn symmetry breaking scale naturally lies on the mirage messenger scale
due to the mixed modulus-anomaly mediation with the gauge coupling unification.
Consequently,  the higgsino mass term is of order the weak scale while the gravitino mass is of PeV order.
This hierarchy naturally leads to the correct electroweak symmetry breaking. 
The higgsinos are thus in the range accessible at future lepton colliders,  while other sparticles
are well-above the current LHC reach and consistent with the observed Higgs boson.
The axino is dominantly produced from the modulus decay and accounts for the correct dark matter abundance.

\end{abstract}

\maketitle 

\section{Introduction}

Supersymmetry (SUSY) is one of the most promising new physics to alleviate 
the quadratic sensitivity of the Higgs sector to unknown ultraviolet (UV) physics
which is genuinely present in the standard model (SM).
The simplest realization of SUSY is the minimal supersymmetric standard model (MSSM)
where all the SM particles own their supersymmetric counterparts.
At the low energy scale,   SUSY is softly broken,   and consequently
the Higgs sector becomes quadratically sensitive only to the sparticle
soft masses.
For this reason,  
weak scale sparticles have been expected to naturally accommodate 
the electroweak symmetry breaking (EWSB),  
and have been considered as one of the main physics targets of the large hadron collider (LHC).

The LHC has, however, increased constraints on the weak scale SUSY,
and the current mass limit for colored sparticles is of the TeV order,   
{\it e.g.},~for simple mSUGRA/CMSSM scenarios~\cite{Aaboud:2017vwy,Sirunyan:2019xwh}.
SUSY may not be solely responsible for the natural EWSB.
Instead, it can be stabilized by another mechanism 
such as the cosmological relaxation of the Higgs boson mass~\cite{Graham:2015cka}.   
Nevertheless,   SUSY is still the best way to protect the Higgs sector against enormous
quantum corrections,     
and at the same time it leads to the successful gauge coupling unification~\cite{Dimopoulos:1981yj}.
Furthermore,  it provides a robust mechanism for dark matter production,  
cosmological inflation,   {\it etc}.

Meanwhile,  the Kachru-Kalosh-Linde-Trivedi (KKLT) setup shows a successful incarnation of de Sitter  
vacua from string theory~\cite{Kachru:2003aw},   
which provides a robust ground for the soft SUSY breaking 
in the MSSM.
In the KKLT, SUSY breaking is generated by an anti-D3 brane, 
and is mediated to the MSSM sector by the model-independent supercoformal 
anomaly~\cite{Randall:1998uk,Giudice:1998xp,Pomarol:1999ie}.
Furthermore,  a sizable SUSY breaking effect is also mediated by $F$-terms of the K\"ahler moduli
which are generated by the anti-D3 brane.
The overall SUSY breaking is realized by the mixed modulus-anomaly mediation~\cite{Choi:2004sx,Choi:2005ge}.
In this setup,  sparticle masses are unified at an intermediate scale, dubbed the mirage messenger scale,
where no physical thresholds are involved.

Although the KKLT setup is attractive, 
it demands a further modification of the Higgs sector because 
it does not automatically generate the higgsino mass and Higgs mixing parameters,
$\mu$ and $B$,  at the correct scales.
In the KKLT,   the sparticles have masses around $m_{3/2}/8\pi^2$ with $m_{3/2}$
being the gravitino mass,   while the Higgs mixing parameter is much larger,  
\begin{equation}
B \sim m_{3/2},
\end{equation}
unless  induced by a super-Weyl invariant operator. 
Such a large $B$ has been considered problematic because the EWSB requires
\begin{equation}
B\mu \sim \frac{ m^2_{3/2} } {(8\pi^2)^2},
\end{equation}
so it necessitates another mechanism to suppress it down to the sparticle
masses.   
 
In this paper,  we point out that the KKLT can naturally induce the correct EWSB 
even for $B$ much larger than the sparticle masses.
The idea is to invoke the Kim-Nilles (KN) mechanism~\cite{Kim:1983dt} 
to generate small $\mu$ from a Planck-suppressed Peccei-Quinn (PQ) symmetric operator.
The mirage mediation {radiatively} generates the PQ scale at the mirage messenger scale $M_{\rm mir}$,  
resulting in
\begin{equation}
\mu \sim \frac{M^2_{\rm mir}}{M_{Pl}} \sim \frac{m_{3/2}}{(8\pi^2)^2},
\end{equation}
{\it insensitively} to the details of the model. 
The only requirement is the proper choice of discrete numbers associated with
the K\"ahler moduli. 
Here $M_{Pl}$ is the reduced Planck mass. 
The above unconventional solution to the $\mu/B\mu$ problem works in high scale 
SUSY with a PeV or heavier gravitino~\cite{Jeong:2011sg} 
because the LEP bound on the chargino mass requires $\mu$ above $104$~GeV~\cite{Zyla:2020zbs}.

High scale SUSY under consideration leaves the higgsinos light,  in the mass range
accessible at future lepton colliders,  while other sparticles appear
around or above $10$~TeV,  well above the LHC limits. 
Such a hierarchy between the higgsinos and other sparticles also allows the
gauge coupling unification as precisely as in the weak scale SUSY case
if the gravitino is below $100$~PeV. 
The gauge coupling unification 
is indeed a prerequisite for the mirage unification of 
sparticle masses~\cite{Jeong:2020wum}.

Our scenario includes two dark matter candidates,  the axion solving the strong CP 
problem~\cite{Peccei:1977hh} and its fermionic partner,  the axino.
For a gravitino mass between PeV and $100$~PeV,  the axion constitutes only a small fraction 
of the dark matter of the universe.
Meanwhile,   as feebly coupled to the MSSM sector via the KN term, axinos are
produced through the freeze-in process.
The freeze-in production should, however, be suppressed to avoid the overclosure of the universe 
because the axino mass is not much below GeV in our scenario. 
This problem is avoided by moduli domination that takes place after 
the primordial inflation unless the moduli are heavier than the inflation scale. 
In the case where the universe experiences moduli mediation,  
the main production process of axino dark matter is the decay of heavy gravitinos produced from 
moduli decay.  
We find that a GeV scale axino successfully accounts for the observed dark matter density.

This paper is organized as follows.  
In section~\ref{sec:MM},   we present a brief review on the mirage unification of 
sparticle masses realized in the KKLT.  
High scale SUSY generally suffers a severer $\mu$-problem due to a large $B$.
We show in section~\ref{sec:PQ-HS} that mirage mediation can radiatively fix the PQ scale
at the mirage messenger scale insenstively to the details of the model,   and then allows
the KN mechanism to resolve the $\mu$ problem in high scale SUSY with a large $B$.
The cosmological aspects of our scenario are described in section~\ref{sec:C}. 
The final section is for conclusions.

\section{\label{sec:MM}Mirage mediation}

In the KKLT,  K\"ahler moduli are stabilized by nonperturbative 
effects in the superpotential while acquiring large supersymmetric masses,  
and their $F$-terms are induced due to SUSY breaking in anti-D3 brane,
{\it i.e.},~after adding the uplifting scalar potential. 
The moduli $F$-terms are loop-suppressed compared to the gravitino mass,
making moduli mediation comparable to anomaly mediation.
Combined with the gauge coupling unification,  
the KKLT leads to mirage mediation,  which effectively corresponds to pure
moduli mediation transmitted at the mirage messenger scale~\cite{Choi:2005uz},
\begin{equation}
M_{\rm mir} = M_{\rm GUT} \left( \frac{m_{3/2}}{M_{Pl}} \right)^{\alpha/2}, 
\end{equation}
with $M_{\rm GUT}$ being the unification scale. 
Here the constant $\alpha$ measures the relative strength of anomaly mediation,
and is a positive rational number determined by the moduli dependence of the K\"ahler
and uplifting potential. 
The original KKLT leads to $\alpha=1$,  which is the case of our interest.
The mirage mediation preserves CP and flavor symmetry, respectively, due to the associated axionic shift symmetries and flavor-universal modular weights. 
These features are phenomenologically important unless the sparticles are very heavy,  
around or above $100$~TeV.

In the mirage mediation,  the gaugino masses at a low energy scale $Q$ are given by  
\begin{equation}
\frac{M_a}{M_0} = 1 - \frac{b_a g^2_a}{4\pi^2} \ln\left(\frac{M_{\rm mir}}{Q} \right),
\end{equation} 
provided that the gauge couplings are universal at $M_{\rm GUT}$. 
Here  $b_a$ are the one-loop beta function coefficients. 
The parameter $\alpha$ is defined by
\begin{equation}
\alpha \equiv \frac{m_{3/2}}{M_0 \ln(M_{Pl}/m_{3/2})}.
\end{equation} 
Interestingly, the gaugino masses unify at  $M_{\rm mir}$, although no physical thresholds appear
at the scale.    
The scalar soft parameters also take the mirage pattern
\begin{eqnarray} 
\label{sparticle-masses}
\frac{m^2_i}{M^2_0} &=&
c_i + 
\frac{1}{4\pi^2} \left\{
\gamma_i - \frac{\dot \gamma_i}{2}\ln\left(\frac{M_{\rm mir}}{Q} \right)
\right\}
\ln\left(\frac{M_{\rm mir}}{Q} \right),
\nonumber \\
\frac{A_{ijk}}{M_0} &=& 1 + \frac{\gamma_i +\gamma_j + \gamma_k}{8\pi^2} 
\ln\left(\frac{M_{\rm mir}}{Q} \right),
\end{eqnarray}
if  $c_i +c_j + c_k = 1$ for those having a Yukawa coupling $y_{ijk}$,
where $c_i$ parameterizes the moduli-mediated contribution and is a rational number 
of order unity depending on the location of the matter in extra dimensions.
Here $\gamma_i$ is the anomalous dimension, 
and the dot denotes differentiation with respect to
$\ln Q$.

The mirage pattern of sparticle masses,
which is in turn essential to solve the $\mu$ problem,  
is a result of the universal gauge couplings at $M_{\rm GUT}$.  
The high scale SUSY can still achieve the gauge coupling unification 
without high scale threshold corrections,  as precisely as the weak scale SUSY,
if the higgsinos are substantially lighter than the other sparticles:
\begin{equation}
\mu \sim 300\,{\rm GeV}
\left( \frac{m_\ast}{{\rm TeV} } \right)^{19/12}
\left(\frac{m_{3/2}}{{\rm PeV}}\right)^{-7/12},
%
\end{equation}
when the other MSSM sparticles are around $m_{3/2}/8\pi^2$. 
Here the scale $m_\ast$,  which determines how precisely the gauge couplings unify  
at $M_{\rm GUT}\sim 10^{16}$~GeV,   should lie in the range between a few hundred GeV and 
$10$~TeV in order to accomplish the 
unification within a few \% deviation~\cite{Krippendorf:2013dqa,Jeong:2017hgz}. 
Combined with the experimental lower bound on the higgsino mass,  
the gauge coupling unification indicates that the gravitino mass is 
below $100$~PeV in the high scale SUSY leaving only the higgsino around the weak scale.

\section{\label{sec:PQ-HS}PQ symmetric Higgs sector}

The minimization condition of the Higgs scalar potential requires 
\begin{equation}
|B\mu| = \frac{\sin2\beta}{2} 
\left( m^2_{H_d} + m^2_{H_u} + 2|\mu|^2 \right), 
\end{equation} 
where the parameters are evaluated near the weak scale,  and 
$\tan\beta$ is the ratio of the vacuum expectation values of the up- and down-type 
Higgs doublets. 
For the case where anomaly mediation is sizable,  the Higgs mixing parameter $B$  
takes a value
\begin{equation}
|B| \sim m_{3/2},
\end{equation}  
unless the higgsino mass parameter $\mu$ is generated from an
operator preserving the super-Weyl symmetry.
This gives rise to a problem
because the scalar soft masses are one-loop suppressed relative to the gravitino
mass.   
On the other hand, the successful EWSB is still possible if the higgsinos are as light as
\begin{equation}
|\mu|  \sim \frac{m^2_{H_d} + m^2_{H_u} }{|B|\tan\beta} \sim \frac{m_{3/2}}{(8\pi^2)^2},
\end{equation}
for moderate $\tan\beta$.
From the LEP bound on the chargino mass, $\mu$ should be larger than $104$~GeV,
so the above scenario with $\mu\sim m_{3/2}/(8\pi^2)^2$ 
is allowed in the high scale SUSY with
\begin{equation}
{\rm PeV} \lesssim m_{3/2} \lesssim 100~{\rm PeV},
\end{equation} 
where the upper bound comes from the condition of gauge coupling unification.

The hierarchy between the higgsino and other sparticle masses in the high scale
SUSY calls for some explanation.  
The mirage mediation naturally addresses the $\mu$ problem if 
one extends the Higgs sector to include the axion superfield $S$ solving the 
strong CP problem.
After the PQ symmetry breaking, the higgsinos receive their mass from the KN type 
superpotential  
\begin{equation}
\Delta W_{\rm KN} = \kappa \frac{S^2}{M_{Pl}}  H_u H_d,
\end{equation}
where $\kappa$ is a constant of order unity. 
The values of $\mu$ and $B$ are determined by how $S$ is stabilized:
\begin{eqnarray}
\label{mu-bmu}
\mu &=& \frac{\kappa}{2} \frac{f^2}{M_{Pl}}, 
\nonumber \\
B &=& 2 \frac{F^S}{S} - m_{3/2},
\end{eqnarray} 
where $f=\sqrt2  \langle |S| \rangle$ is the axion decay constant,
and $F^S$ is the $F$-term of $S$.  
The dependence of $B$ on the gravitino mass reflects the fact that
the KN term,  which is an operator of mass dimension $4$,  explicitly
breaks the conformal symmetry that is a part of the super-Weyl symmetry.

To stabilize the saxion,  we consider a PQ symmetric Yukawa interaction  
\begin{equation}
\Delta W_{\rm PQ} = y S \Phi \Phi^c,
\end{equation}  
where the PQ messengers $\Phi+\Phi^c$ belong to  $5+\bar 5$ representation under SU$(5)$ into which 
the SM gauge groups are embedded.
The saxion potential is generated by integrating out the heavy messengers in 
the large background value of the saxion~\cite{Pomarol:1999ie}
\begin{equation}
V \simeq m^2_S |S|^2,
\end{equation}
where $m^2_S$ is the soft scalar mass squared of the saxion 
renormalized at the scale,  $Q=|S|$.
Taking the modular weights of $S$ and $\Phi+\Phi^c$ to 
be $c_S = 0$ and $c_\Phi=c_{\Phi^c}=1/2$ so as to satisfy the mirage conditions,  
one finds  
\begin{equation}
\label{ms}
\frac{m^2_S}{M^2_0}  =\frac{1}{8\pi^2} 
\left( \gamma_S
 - \dot \gamma_S \ln \left( \frac{M^2_{\rm mir}}{|S|^2} \right)
\right)
\ln \left( \frac{M^2_{\rm mir}}{|S|^2} \right),
\end{equation}
with $\gamma_S= -y^2$, 
as follows from the relation (\ref{sparticle-masses}).
The minimum of the potential appears around the saxion value at which $m^2_S$ crosses zero, 
{\it i.e.},~at $|S| \approx M_{\rm mir}$, 
implying that the axion decay constant is fixed at
\begin{equation}
\label{decay-constant}
f \approx M_{\rm GUT} \left( \frac{m_{3/2}}{M_{Pl}} \right)^{\frac{1}{2}}
\sim 
10^{10}{\rm GeV} \left( \frac{m_{3/2}}{{\rm PeV}} \right)^{\frac{1}{2}},
\end{equation} 
for $\alpha=1$.  
The $F^S/S$ is $\sim m_{3/2}/(8\pi^2)$
at the potential minimum. 
Therefore,  one finds
\begin{eqnarray}
\mu &\approx& \frac{\kappa}{2} \left(\frac{M_{\rm GUT}}{M_{Pl}}\right)^2 m_{3/2}
\sim \frac{m_{3/2}}{(8\pi^2)^2},
\nonumber \\
B &\simeq& -m_{3/2},
\end{eqnarray}
where we have used the fact that the ratio between $M_{\rm GUT}$ and $M_{Pl}$ is numerically
close to a loop factor. 
It is a remarkable that the radiatively stabilized saxion naturally generates $\mu$ and $B$
at the right scales as required for the EWSB,  
insensitively to the details of the model.
This relies only on the
proper choice of discrete numbers,  $\alpha$ and $c_i$,
associated with the K\"ahler moduli.

It is  straightforward to see that the saxion $\sigma$ and axino $\tilde a$ acquire masses 
respectively according to
\begin{eqnarray}
m_\sigma &=& 
\sqrt{ \frac{5y^2 }{4\pi^2} } M_0
\sim   y \frac{m_{3/2}}{(8\pi^2)^{3/2}},
\nonumber \\
m_{\tilde a} &=& \frac{5 y^2}{16\pi^2} M_0
\sim y^2 \frac{m_{3/2}}{(8\pi^2)^2}.
\end{eqnarray}   
The axino is much lighter than the higgsinos for 
a small Yukawa coupling.  
On the other hand,
the axion becomes massive due to its anomalous coupling to gluons
while dynamically cancelling off the strong CP phase.

After the EWSB,  the $B\mu$ term makes the saxion mix with the SM-like neutral Higgs boson $h$.
For the case where the saxion mass is below the weak scale, 
the mixing angle is roughly estimated by
\begin{equation}
\theta_{\rm mix} \sim \frac{B\mu}{m^2_h \tan\beta}
\frac{v}{f}  \sim \frac{10^{-4}}{\tan\beta}  \left(
\frac{m_{3/2}}{\rm PeV} \right)^{3/2},
\end{equation}
where the last estimation follows from
the relations (\ref{mu-bmu}) and (\ref{decay-constant}).
The mixing angle is smaller than about $0.1$ for $m_{3/2} \lesssim 100$~PeV.
The saxion decays into SM particles through the mixing if  kinematically
allowed.   
It is also straightforward to see that the KN term  induces the Yukawa interaction
\begin{equation}
\label{higgsino-axino}
\lambda h \tilde h \tilde a,
\end{equation}
with the coupling constant given by
\begin{equation}
\lambda \sim \frac{\mu}{f} \sim 10^{-8} 
\left(
\frac{m_{3/2}}{\rm PeV} \right)^{1/2}, 
\end{equation}
where $\tilde a$ is the axino,  and $\tilde h$ is the neutral higgsino.
The above interaction determines the decay rate of the heavier of $\tilde h$ and  $\tilde a$.

We close this section by mentioning that  the PQ sector properties presented above hold 
for $y^2 \gtrsim 1/8\pi^2$ since otherwise $m^2_S$ receives a sizable
contribution from higher loops of gauge-charged heavy PQ messengers. 
If one considers a superpotential term $y^\prime SNN$ instead of $y S\Phi\Phi^c$  with 
$N$ being the gauge singlet,   there are no sizable contributions from higher loops of $N$.
However,  the radiative stabilization at at the scale 
$f\approx M_{\rm mir}$ still requires  
$y^{\prime 2}\gtrsim 1/8\pi^2$
because
$c_S$ receives small quantum corrections of the order of $1/8\pi^2$
from nonperturbative effects,
string loops,  and higher order $\alpha^\prime$ 
corrections~\cite{Becker:2002nn,Berg:2005ja,Anguelova:2010ed}. 
An intriguing aspect of the model with
$y^\prime SNN$ is that  the baryon asymmetry
of the universe can be explained  via the Affleck-Dine leptogenesis~\cite{Affleck:1984fy} 
realized along the $LH_u$ flat direction~\cite{Bae:2016zym}.

\section{\label{sec:C}Cosmology}

In our scenario, both the axion and the lightest sparticle can contribute to the dark matter. 
For the case where PQ symmetry is broken during the primordial inflation and never
restored afterwards, the axion abundance from the misalignment reads
\begin{equation}
\Omega_a h^2 \simeq
0.1\times 10^{-2}\,\theta^2_{\rm mis} 
\left( \frac{f}{10^{10}{\rm GeV}} \right)^{1.19},
\end{equation} 
neglecting the anharmonic effect~\cite{Abbott:1982af,Preskill:1982cy,Dine:1982ah,Turner:1985si,Bae:2008ue,Visinelli:2009zm}.
Here $\theta_{\rm mis}\equiv a_{\rm mis}/f$ is the initial  angle of the axion.
For $f\lesssim 10^{11}$~GeV, 
the axion abundance is the only small part of the total dark matter abundance
unless it lies very close to the hilltop of the potential 
so that the anharmonic effect is significant.
Hence, the axion is the subdominant component of dark matter in our case.

The production of the lightest sparticle depends on the cosmological evolution 
of the moduli and saxion. 
The string moduli generally have an initial condensation of order $M_{Pl}$ after the inflation,  
and then their coherent oscillations soon dominate the energy density 
of the universe.
As a consequence, the moduli-induced gravitino problem arises in the KKLT
because the gravitinos from modulus decays produce too many neutralinos,  overclosing
the universe if the lightest neutralino is stable~\cite{Endo:2006zj,Nakamura:2006uc,Dine:2006ii}.

Our scenario predicts the neutral higgsino as the lightest sparticle in the MSSM sector,
but the axino is lighter if the Yukawa coupling responsible 
for the saxion stabilization is less than one.   
Feebly coupled to the Higgs sector through the interaction (\ref{higgsino-axino}),
axinos are produced by  the freeze-in mechanism. 
This process, however, produces too many axinos unless the axino mass is of keV order 
or smaller~\cite{Bae:2017dpt}, 
or the reheating temperature is very low. 
One may  rely on moduli domination,  which occurs if some moduli are lighter
than the inflation scale.   
The axino density is diluted if the modulus decays follow the axino freeze-in 
process.    
This case, however confronts another difficulty 
because axinos are non-thermally produced from subsequent decays of moduli and gravitino. 
In the original KKLT,  where the K\"ahler moduli are much heavier than the gravitino,   
the freeze-in process and modulus decays causes overproduction of the axino 
unless the axino mass is much below GeV~\cite{Nakamura:2008ey,Choi:2009qd}.
It is difficult to accommodate such a light axino in high scale SUSY
because the radiative saxion stabilization puts a lower bound on its mass.

To resolve the axino overproduction problem,   we consider a generalized KKLT 
where all the K\"ahler moduli except one are stabilized by the nonperturbative superpotential 
as in the original KKLT while the remaining one is stabilized through the K\"ahler mixing 
with the others~\cite{Choi:2006za}.
The modulus $F$-terms are universal due to the no-scale structure of 
the K\"ahler potential, leading to the mirage mediation in the same way as the original KKLT.
Another important feature is that the radial component of the lightest modulus 
is relatively light  
\begin{equation}
m_{\phi_1} \simeq \sqrt2 \,m_{3/2},
\end{equation}
independently of the details of the model.\footnote{
The phase component of the lightest modulus has a large decay constant 
around $M_{Pl}/8\pi^2$, and can play the role of the QCD axion~\cite{Choi:2006za}.
In our scenario,   the QCD axion comes mainly from the phase component of the PQ
scalar  because $f\approx M_{\rm mir}$ is much lower than the GUT scale.
The phase component of the lightest modulus acquires a mass from nonperturbative
effects which break the associated shift symmetry. It can make up a sizable fraction of 
the dark matter depending on the model.
}
The modulus $\phi_1$  decays dominantly into the gauge sector with
the  rate  
\begin{equation}
\Gamma_{\phi_1} \sim \frac{3}{32\pi} \frac{m^3_{\phi_1} } {M^2_{Pl}}, 
\end{equation} 
but its decay into gravitinos is kinematically forbidden.
The modulus decay temperature is estimated to be 
\begin{equation}
T_{\phi_1} = \left( \frac{90}{\pi^2 g_\ast} \right)^{ \frac{1}{4}} \sqrt{ \Gamma_{\phi_1} M_{Pl} }
\sim 0.2 \,{\rm GeV}
\left( \frac{m_{\phi_1}}{2 {\rm PeV} }\right)^{\frac{3}{2}},
\end{equation}
where $g_\ast$ counts the number of relativistic degrees of freedom at $T_{\phi_1}$.

As the next-to-lightest K\"ahler modulus $\phi_2$ is stabilized by a nonperturbative superpotential term, 
it acquires a large supersymmetric mass
\begin{equation}
m_{\phi_2} \sim m_{3/2} \ln(M_{Pl}/m_{3/2}),
\end{equation}
and its branching fraction into gravitinos is about
$0.03$~\cite{Endo:2006zj,Nakamura:2006uc,Dine:2006ii}.
Produced gravitinos decay while producing MSSM sparticles that cascade 
to neutral higgsinos. 
The gravitino decay temperature is slightly below $T_{\phi_1}$ because
its decay width is given by $\Gamma_{3/2} \sim 0.1m^3_{3/2}/M^2_{Pl}$.   
The higgsino scattering rate is much larger than its decay rate,
and thus higginos become non-relativistic before their decay.
For the case with $\mu > m_h + m_{\tilde a}$,  
the higgsino decay occurs mainly via the interaction (\ref{higgsino-axino}),
much more effectively than its annihilation at the gravitino decay temperature~\cite{Nakamura:2008ey}.
The axino density from the higgsino decay is diluted due to late time
entropy released from the $\phi_1$ decay. 
The dilution factor is determined by the ratio between moduli decay temperatures
\begin{equation}
\Delta_\phi =
\frac{T_{\phi_2}}{T_{\phi_1} }
\simeq 
\left( \frac{ m_{\phi_2} }{m_{\phi_1} } \right)^{3/2}
\sim 90,
\end{equation}
under the assumption that both $\phi_1$ and $\phi_2$ initially have 
similar condensations.
Finally, the axino abundance reads  
\begin{equation}
\Omega_{\tilde a} h^2 
\sim
0.1
\left(\frac{m_{\tilde a}}{2 {\rm GeV}} \right)
\left( \frac{m_{3/2} }{{\rm PeV}}\right)^{1/2}.
\end{equation}
Thus, for an axino with mass around GeV,   
the moduli domination does not cause the moduli-induced gravitino problem.

It is worth examining the axino abundance produced via the freeze-in process. 
For the gravitino mass around PeV,  the axino freeze-in production takes place most efficiently 
while the coherent oscillation of $\phi_2$ dominates the energy density of the universe. 
See the appendix for more discussion on the freeze-in production during a matter dominated era.
Including the dilution factor from the $\phi_1$ decay,  one finds the axino freeze-in abundance to be
\begin{eqnarray}
\hspace{-0.5cm}
\left.
\Omega_{\tilde a} h^2\right|_{\rm freeze}
&\sim& 
\frac{10^2}{\Delta_\phi}
\left( \frac{\lambda}{10^{-8}} \right)^2
\frac{m_{\tilde a}}{T_{\phi_2}}
\nonumber \\
&\sim&
0.04\left( \frac{\lambda}{10^{-8}} \right)^2
\left( \frac{m_{\tilde a}}{2 {\rm GeV} } \right)
\left( \frac{ m_{3/2} }{ {\rm PeV} } \right)^{-\frac{3}{2}},
\end{eqnarray}
which is smaller than that from the $\phi_2$ decay,
and gets suppressed further for a larger gravitino mass.    
Here we have considered the case where the $\phi_1$ decays well after
the axion freeze-in production completes.

 \begin{figure}[t] \centering
 \includegraphics[width=0.45\textwidth]{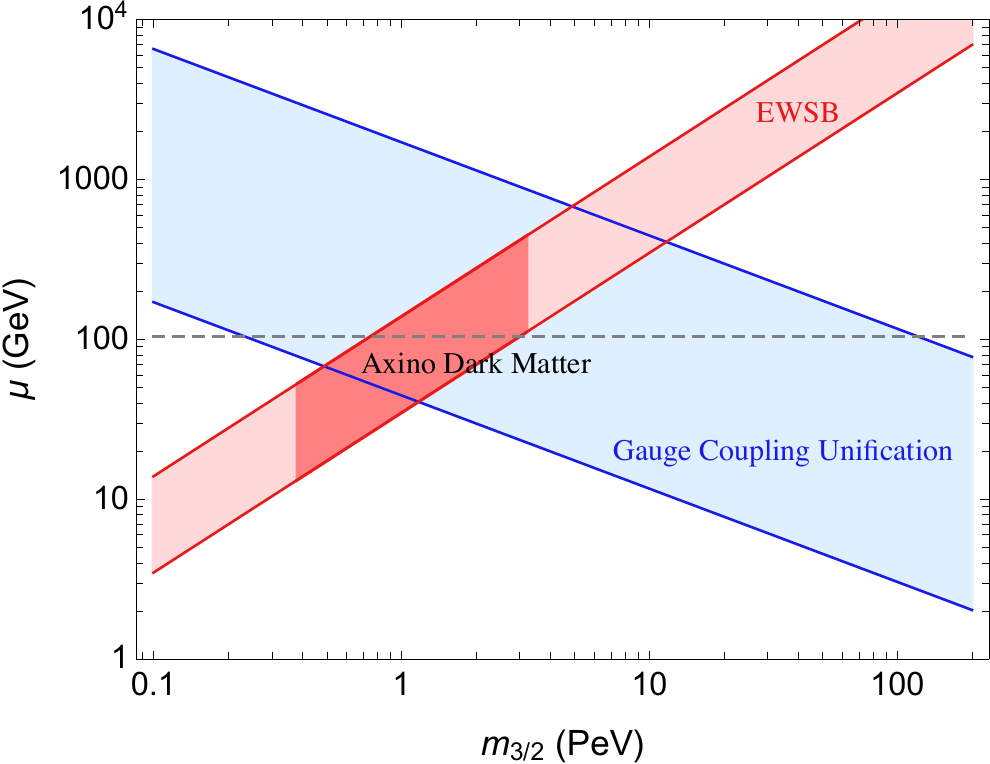}
 \hspace{0.2cm}
 \caption{
Viable region of $\mu$ and $m_{3/2}$ 
in high scale SUSY where sparticles other than 
the higgsinos are around $m_{3/2}/8\pi^2$.
The gauge coupling unification,  which is required for the sparticles to take the mirage pattern,
is achieved in the light blue shaded region. 
For a large $B$ around $m_{3/2}$,  mirage mediation allows 
the KN mechanism to naturally induce $\mu$ at the right scale for the correct EWSB 
as is shown in the light red shaded region.
For the case where the universe experiences moduli domination, 
the axino can explain the dark matter of the universe in the red shaded region,
where we have used that its mass lies in the range 
between about $10^{-6}$ and $10^{-5}$ times $m_{3/2}$.
Meanwhile,  the LEP bound on the chargino mass requires $\mu$  larger than $104$~GeV
as indicated by the gray dashed line. 
 }
 \label{fig:mu}
 \end{figure}

Figure~\ref{fig:mu} shows the viable region of $\mu$ and $m_{3/2}$ 
in the high scale SUSY realized in the KKLT.
A $B$-term around $m_{3/2}$ successfully leads to the EWSB because the KN mechanism generates
$\mu$ at the right scale.
The gauge coupling unification is also achieved by the light higgsinos, allowing the sparticles
to take the mirage pattern. 
Here we have taken $m_\ast$ between $0.3$ and $3$~TeV for the precise gauge coupling unification,
and $\kappa$ between $0.5$ and $2$ for the KN term inducing $\mu$. 
The dark matter can also be explained well by the light axino if moduli domination
takes place. 
As shown in the figure, the EWSB, unification, and dark matter indicate that the gravitino
is around a few PeV, the higginos are around a few hundred GeV, and the axino around
GeV.

We have, up to this point, ignored the saxion coherent oscillations, which 
can dominate the energy density of the universe if the saxion
has a large initial amplitude of the order of $M_{Pl}$, or if
the saxion is trapped at the origin during inflation due to the Hubble-induced mass term.
In the latter case,   the saxion potential energy drives a thermal inflation until the temperature drops 
below the SUSY breaking scale~\cite{Choi:2009qd}.\footnote{
In this case,  PQ symmetry is restored during inflation,  and thus 
the PQ sector should be arranged to have the domain-wall number equal to unity 
to avoid the domain-wall problem.  
}
Note that  the saxion cannot be much lighter than the higgsinos  because the Yukawa coupling
responsible for the saxion stabilization is bounded from below.
As a result,  even if the saxion domination occurs, 
its decay occurs well before the $\phi_2$ decay,   and does not change our results.

Finally, we discuss collider signatures of our scenario. 
Because the higgsinos are much lighter than other MSSM sparticles,
the lightest neutralino and the lightest chargino are almost higgsino-like.
The mass difference between them reads
\begin{equation}
\Delta m \equiv m_{\chi^+_1} - m_{\chi^0_1} 
= \Delta m_{\rm tree} + \Delta m_{\rm loop},
\end{equation}
where the tree-level contribution comes from mixing with the bino and wino
\begin{equation}
\Delta m_{\rm tree} 
\simeq 300\,{\rm MeV}
 \left( \frac{10^4{\rm GeV}}{M_2} \right) 
\left( 1+ 0.3 \frac{M_2}{M_1} \right),
\end{equation}
while the loop contribution is dominantly from gauge boson loops~\cite{Thomas:1998wy} 
\begin{equation}
\Delta m_{\rm loop} \approx
300\,{\rm MeV}
\left( \frac{|\mu|}{300{\rm GeV}} \right)^{0.15}.
\end{equation} 
Here $M_1$ and $M_2$ are the bino and wino mass, respectively, and both are positive
in the mirage mediation with $\alpha=1$.
For $\Delta m$ larger than the pion mass, the lightest chargino decays dominantly 
to the lightest neutralino and the charged pion.
The decay time of the lightest chargino is roughly 
\begin{equation}
\tau_{\chi^+_1} \sim 0.3\times 10^{-11}{\rm sec} 
\left( \frac{\Delta m}{600{\rm MeV}} \right)^{-3}, 
\end{equation} 
and so it would be quite difficult to probe it at the LHC. 
The lightest neutralinos may show a peculiar signature of the long-lived particle  at the LHC
for the coupling $h \tilde h \tilde a$ to be around $10^{-8}$~\cite{Bae:2020dwf}. 
Future lepton colliders are able to probe clearer signals from 
the processes, 
$e^+e^- \to \gamma \chi^0_1 \chi^0_2$,   or $\gamma \chi^+_1 \chi^-_1$,
mediated by a virtual $Z$ boson or photon~\cite{Baer:2011ec,Berggren:2013vfa}.

\section{\label{sec:Con}Conclusions}

As the first explicit realization of 4-dimensional de-Sitter vacua 
with all string moduli stabilized, 
the KKLT provides a natural and interesting framework to realize high scale SUSY where 
the EWSB is successfully 
achieved by the higgsinos much lighter than the other sparticles. 
The EWSB requires $\mu\sim m_{3/2}/(8\pi^2)^2$ if $B$ is 
around the gravitino mass as is generally the case when anomaly mediation is sizable.
For the PQ extended Higgs sector, 
the Kim-Nilles mechanism naturally generates $\mu$ at the right scale,
{\it insensitively} to the details of the model,   if the PQ scale is {\it radiatively} 
fixed at the mirage messenger scale. 
All those features indicate the high scale SUSY with a PeV gravitino 
where the higgsinos are around the weak scale,  thus 
accessible at future lepton colliders,  and  other sparticles are around $10$~TeV
as required to accommodate the $125$~GeV Higgs boson while avoiding the experimental
constraints. 
The modulus domination suppresses the freeze-in axino production, which otherwise leads 
to the overclosure of the universe, 
but it is followed by non-thermal axino production from heavy gravitinos at moduli decay.
The dark matter of the universe can then be well explained by a GeV axino. 
 \\

\noindent{\bf Acknowledgments}
\\
This work was supported by the National Research Foundation of Korea (NRF) grant funded by the Korean government, NRF-2020R1C1C1012452 (KJB) and NRF-2018R1C1B6006061 (KSJ).

\appendix

\section{Freeze-in production of axino}

In the appendix, 
we briefly review the freeze-in axino production from heavy particle decays.
In circumstances where moduli dominate the energy density of the universe or decay,   
the procedure of axino production may differ from 
the standard calculation done in the radiation-dominated regime.

The freeze-in production of particle $X$ from the 
process $A\to B+X$ is described by the following Boltzmann equation 
when $A$ and $B$ are in thermal equilibrium:
\begin{eqnarray}
\frac{dn_X}{dt}+3Hn_X=
\frac{g_Am_A^2\Gamma_A}{2\pi^2}T K_1(m_A/T),
\label{eq:boltz}
\end{eqnarray}
where $g_A$ is the degrees of freedom of $A$, 
$\Gamma_A$ is the partial decay width of $A\to B+X$,
and $K_1$ is the first modified Bessel function of the second kind.

In the radiation-dominated (RD) era, 
one can rewrite the reation~\eqref{eq:boltz} as follows:
\begin{eqnarray}
\frac{H(m_A)s(m_A)}{x^4}  \frac{dY_X}{dx}=\frac{g_Am_A^3\Gamma_A}{2\pi^2x}K_2(x),
\end{eqnarray}
for $x\equiv m_A/T$, 
where $Y_X=n_X/s$ is the yield of $X$, 
$s(T)=(2\pi^2/45)g_{*s}T^3$ is the entropy density, 
and $H(T)=(\pi^2g_*/90)^{1/2}T^2/M_{Pl}$ is the Hubble parameter.
Here,   $g_{*s}$ and $g_*$ are respectively the effective degrees of freedom for the entropy density 
and radiation density,
and we have assumed that they have the same value and are 
constant during freeze-in production.
Hence the yield is obtained by integrating the above equation
\begin{eqnarray}
Y_X(x)=\frac{g_Am_A^3\Gamma_A }{2\pi^2H(m_A)s(m_A)}\int^x dx'~ x'^3K_1(x'),
\end{eqnarray}
or equivalently the number density is given by
\begin{eqnarray}
n_X(x)=\frac{g_Am_A^3\Gamma_A }{2\pi^2H(m_A)}x^{-3}\int^x dx'~ x'^3K_1(x').
\end{eqnarray}

In the decaying-particle-dominated (DD) era,   the entropy is not conserved,  and thus 
one needs to deal with the total number of $X$, 
$N_X$,   instead of the number density.
From \eqref{eq:boltz},   one can find
\begin{eqnarray}
\frac{1}{R(T)^3}\frac{dN_X}{dt}=\frac{g_Am_A^2\Gamma_A}{2\pi^2}T K_1(m_A/T),
\label{eq:boltz2}
\end{eqnarray}
where the scale factor is given by $R(T)=R(m_A)x^{8/3}$. 
In the meantime,   the Hubble parameter is written
\begin{eqnarray}
H(T)=\left(\frac{\pi^2g_*}{90}\right)^{1/2}\frac{T^4}{T_D^2M_P}=H(m_A)x^{-4},
\end{eqnarray}
where $T_D$ is the decay temperature of the dominating particle.
Hence the relation~\eqref{eq:boltz2} leads to
\begin{eqnarray}
\frac{dN_X}{dx}=\frac{8R(m_A)^3}{3H(m_A)}\frac{g_Am_A^3\Gamma_A}{2\pi^2}x^{10}K_1(x), 
\end{eqnarray}
implying that the number density is given by 
\begin{eqnarray}
n_X(x)&=&\frac{8}{3H(m_A)}\frac{R(m_B)^3}{R(T)^3}\frac{g_Am_A^3\Gamma_A}{2\pi^2}\nonumber\\
&&\times\int^x dx'~x'^{10}K_1(x') \, .
\end{eqnarray}
For the case with $m_A=300$~GeV and $T_{\phi}=10$~GeV,  one finds 
\begin{eqnarray}
\frac{R(m_A)^3}{R(T_{\phi})^3}\simeq\frac{T_{\phi}^8}{m_A^8}\sim10^{-12}\, ,
\label{eq:dilut}
\end{eqnarray}
and
\begin{eqnarray}
\int^{10} dx'~x'^{10}K_1(x')\simeq 10^{6}\, .
\end{eqnarray}
Therefore,  the number density reads
\begin{eqnarray}
n_X(T_{\phi})&\sim& 10^{-6}\frac{8g_A m_A^3}{6\pi^2}\frac{\lambda^2m_A}{8\pi}
\left(\frac{90}{\pi^2g_*}\right)^{1/2}\frac{T_{\phi}^2M_P}{m_A^4}\nonumber\\
&\sim&10^{-3}~\text{GeV}^3\times \left(\frac{\lambda}{10^{-8}}\right)^2\left(\frac{T_{\phi}}{10~\text{GeV}}\right)^2\, .
\end{eqnarray}
Meanwhile,  because the entropy at that time is given by
\begin{eqnarray}
\hspace{-0.8cm}
s(T_{\phi})=\frac{2\pi^2}{45}g_{*s}T_{\phi}^3\sim 10^5~\text{GeV}^3\times \left(\frac{T_{\phi}}{10~\text{GeV}}\right)^3,
\end{eqnarray}
the yield is found to be 
\begin{eqnarray}
Y_X(T_{\phi})&=&\frac{n_X(T_{\phi})}{s(T_{\phi})}\nonumber\\
&\sim& 10^{-8}\times \left(\frac{\lambda}{10^{-8}}\right)^2\left(\frac{10~\text{GeV}}{T_{\phi}}\right)\, .
\end{eqnarray}
If no entropy production occurs after the freeze-in production, 
the relic density  of $X$ becomes
\begin{eqnarray}
\Omega_X h^2&\simeq& 2.8\times 10^8 Y_X\left(\frac{m_X}{\text{GeV}}\right)\nonumber\\
&\sim&10\times \left(\frac{\lambda}{10^{-8}}\right)^2\left(\frac{10~\text{GeV}}{T_{\phi}}\right)\left(\frac{m_X}{\text{GeV}}\right), 
\end{eqnarray}
assuming the relation~\eqref{eq:dilut}.



\begin{thebibliography}{99}

 

\bibitem{Aaboud:2017vwy}
M.~Aaboud \textit{et al.} [ATLAS],
Phys. Rev. D \textbf{97}, no.11, 112001 (2018)
[arXiv:1712.02332 [hep-ex]].

\bibitem{Sirunyan:2019xwh}
A.~M.~Sirunyan \textit{et al.} [CMS],
Eur. Phys. J. C \textbf{80}, no.1, 3 (2020)
[arXiv:1909.03460 [hep-ex]].




\bibitem{Graham:2015cka}
P.~W.~Graham, D.~E.~Kaplan and S.~Rajendran,
Phys. Rev. Lett. \textbf{115}, no.22, 221801 (2015)
[arXiv:1504.07551 [hep-ph]].



\bibitem{Dimopoulos:1981yj}
S.~Dimopoulos, S.~Raby and F.~Wilczek,
Phys. Rev. D \textbf{24} (1981), 1681-1683.


\bibitem{Kachru:2003aw} 
  S.~Kachru, R.~Kallosh, A.~D.~Linde and S.~P.~Trivedi,
  Phys.\ Rev.\ D {\bf 68}, 046005 (2003)
  [hep-th/0301240].
  
\bibitem{Randall:1998uk} 
  L.~Randall and R.~Sundrum,
  Nucl.\ Phys.\ B {\bf 557}, 79 (1999)
  [hep-th/9810155].

\bibitem{Giudice:1998xp} 
  G.~F.~Giudice, M.~A.~Luty, H.~Murayama and R.~Rattazzi,
  JHEP {\bf 9812}, 027 (1998)
  [hep-ph/9810442].


\bibitem{Pomarol:1999ie} 
  A.~Pomarol and R.~Rattazzi,
  JHEP {\bf 9905}, 013 (1999)
  [hep-ph/9903448].



\bibitem{Choi:2004sx}
K.~Choi, A.~Falkowski, H.~P.~Nilles, M.~Olechowski and S.~Pokorski,
JHEP \textbf{11}, 076 (2004)
[arXiv:hep-th/0411066 [hep-th]].

\bibitem{Choi:2005ge}
K.~Choi, A.~Falkowski, H.~P.~Nilles and M.~Olechowski,
Nucl. Phys. B \textbf{718}, 113-133 (2005)
[arXiv:hep-th/0503216 [hep-th]].


\bibitem{Kim:1983dt}
J.~E.~Kim and H.~P.~Nilles,
Phys. Lett. B \textbf{138}, 150-154 (1984).


\bibitem{Jeong:2011sg} 
  K.~S.~Jeong, M.~Shimosuka and M.~Yamaguchi,
  JHEP {\bf 1209}, 050 (2012)
  [arXiv:1112.5293 [hep-ph]].


\bibitem{Zyla:2020zbs} 
  P.~A.~Zyla {\it et al.} [Particle Data Group],
  PTEP {\bf 2020}, no. 8, 083C01 (2020).
 



\bibitem{Jeong:2020wum}
K.~S.~Jeong and C.~B.~Park,
[arXiv:2011.11993 [hep-ph]].

 
 
\bibitem{Peccei:1977hh} 
  R.~D.~Peccei and H.~R.~Quinn,
  Phys.\ Rev.\ Lett.\  {\bf 38}, 1440 (1977);
%
  Phys.\ Rev.\ D {\bf 16}, 1791 (1977).
 


\bibitem{Choi:2005uz} 
  K.~Choi, K.~S.~Jeong and K.~i.~Okumura,
  JHEP {\bf 0509}, 039 (2005)
 [hep-ph/0504037].

 
\bibitem{Krippendorf:2013dqa} 
  S.~Krippendorf, H.~P.~Nilles, M.~Ratz and M.~W.~Winkler,
  Phys.\ Rev.\ D {\bf 88}, 035022 (2013)
  [arXiv:1306.0574 [hep-ph]].
 
 
\bibitem{Jeong:2017hgz} 
  K.~S.~Jeong,
  Phys.\ Lett.\ B {\bf 769}, 42 (2017)
  [arXiv:1701.06947 [hep-ph]].
 

  
\bibitem{Becker:2002nn} 
  K.~Becker, M.~Becker, M.~Haack and J.~Louis,
  JHEP {\bf 0206}, 060 (2002)
  [hep-th/0204254].
  
\bibitem{Berg:2005ja} 
  M.~Berg, M.~Haack and B.~Kors,
  JHEP {\bf 0511}, 030 (2005)
  [hep-th/0508043].
 
\bibitem{Anguelova:2010ed} 
  L.~Anguelova, C.~Quigley and S.~Sethi,
  JHEP {\bf 1010}, 065 (2010)
  [arXiv:1007.4793 [hep-th]].
 
 
\bibitem{Affleck:1984fy} 
  I.~Affleck and M.~Dine,
  Nucl.\ Phys.\ B {\bf 249}, 361 (1985).
  
  
\bibitem{Bae:2016zym} 
  K.~J.~Bae, H.~Baer, K.~Hamaguchi and K.~Nakayama,
  JHEP {\bf 1702}, 017 (2017)
  [arXiv:1612.02511 [hep-ph]].

\bibitem{Abbott:1982af}
L.~F.~Abbott and P.~Sikivie,
Phys. Lett. B \textbf{120}, 133-136 (1983)

\bibitem{Preskill:1982cy}
J.~Preskill, M.~B.~Wise and F.~Wilczek,
Phys. Lett. B \textbf{120}, 127-132 (1983)

\bibitem{Dine:1982ah}
M.~Dine and W.~Fischler,
Phys. Lett. B \textbf{120}, 137-141 (1983)

\bibitem{Turner:1985si}
M.~S.~Turner,
Phys. Rev. D \textbf{33}, 889-896 (1986)

\bibitem{Bae:2008ue}
K.~J.~Bae, J.~H.~Huh and J.~E.~Kim,
JCAP \textbf{09}, 005 (2008)
[arXiv:0806.0497 [hep-ph]].

\bibitem{Visinelli:2009zm} 
  L.~Visinelli and P.~Gondolo,
  Phys.\ Rev.\ D {\bf 80}, 035024 (2009)
  [arXiv:0903.4377 [astro-ph.CO]].



\bibitem{Endo:2006zj} 
  M.~Endo, K.~Hamaguchi and F.~Takahashi,
  Phys.\ Rev.\ Lett.\  {\bf 96}, 211301 (2006)
  [hep-ph/0602061].
  
\bibitem{Nakamura:2006uc} 
  S.~Nakamura and M.~Yamaguchi,
  Phys.\ Lett.\ B {\bf 638}, 389 (2006)
  [hep-ph/0602081].
  
\bibitem{Dine:2006ii} 
  M.~Dine, R.~Kitano, A.~Morisse and Y.~Shirman,
  Phys.\ Rev.\ D {\bf 73}, 123518 (2006)
  [hep-ph/0604140].
  
  
 
\bibitem{Bae:2017dpt} 
  K.~J.~Bae, A.~Kamada, S.~P.~Liew and K.~Yanagi,
  JCAP {\bf 1801}, 054 (2018)
  [arXiv:1707.06418 [hep-ph]].
   
  
  
  
\bibitem{Nakamura:2008ey} 
  S.~Nakamura, K.~i.~Okumura and M.~Yamaguchi,
  Phys.\ Rev.\ D {\bf 77}, 115027 (2008)
  [arXiv:0803.3725 [hep-ph]].


\bibitem{Choi:2009qd} 
  K.~Choi, K.~S.~Jeong, W.~I.~Park and C.~S.~Shin,
  JCAP {\bf 0911}, 018 (2009)
  [arXiv:0908.2154 [hep-ph]].



\bibitem{Choi:2006za} 
  K.~Choi and K.~S.~Jeong,
  JHEP {\bf 0701}, 103 (2007)
  [hep-th/0611279].



\bibitem{Thomas:1998wy} 
  S.~D.~Thomas and J.~D.~Wells,
  Phys.\ Rev.\ Lett.\  {\bf 81}, 34 (1998)
  [hep-ph/9804359].

\bibitem{Bae:2020dwf}
K.~J.~Bae, M.~Park and M.~Zhang,
Phys. Rev. D \textbf{101}, no.11, 115036 (2020)
[arXiv:2001.02142 [hep-ph]].

\bibitem{Baer:2011ec}
H.~Baer, V.~Barger and P.~Huang,
JHEP \textbf{11}, 031 (2011)
[arXiv:1107.5581 [hep-ph]].


\bibitem{Berggren:2013vfa}
M.~Berggren, F.~Br\"ummer, J.~List, G.~Moortgat-Pick, T.~Robens, K.~Rolbiecki and H.~Sert,
Eur. Phys. J. C \textbf{73}, no.12, 2660 (2013)
[arXiv:1307.3566 [hep-ph]].



 
 
 




  
\end{thebibliography}
\end{document}